\documentclass[a4paper,preprint]{raa}

\usepackage{graphicx,times,natbib}
\usepackage[T1]{fontenc}
\usepackage{fancyvrb}

\newcommand{\psrchive}{{\sc psrchive}}
\newcommand{\psr}{PSR~J0437$-$4715}

\newcommand{\backtick}{\symbol{0}}

\newcommand{\On}[1]{\ensuremath{#1_{\rm on}}}
\newcommand{\Off}[1]{\ensuremath{#1_{\rm off}}}

\newcommand{\Ssys}{\ensuremath{S_{\rm sys}}}

\begin{document}

\title{Pulsar data analysis with \psrchive}

\setcounter{page}{1}

    \author{Willem van Straten 
       \inst{1}
    \and Paul Demorest
       \inst{2}
    \and Stefan Os{\l}owski
       \inst{1,3}
    }

    \institute{Centre for Astrophysics \& Supercomputing,
               Swinburne University of Technology, \\
               PO Box 218, Hawthorn VIC 3122, Australia;
               {\it willem@swin.edu.au} \\
           \and
               National Radio Astronomy Observatory,
               520 Edgemont Rd., Charlottesville, VA 22903, USA \\
           \and
               CSIRO Astronomy and Space Sciences, PO Box 76, Epping, NSW 1710, Australia}

\abstract{ \psrchive\ is an open-source, object-oriented,
     scientific data analysis software library and application suite
     for pulsar astronomy.
     It implements an extensive range of general-purpose algorithms
     for use in data calibration and integration, statistical analysis
     and modeling, and visualisation.
     These are utilised by a variety of applications specialised for
     tasks such as pulsar timing, polarimetry, radio frequency
     interference mitigation, and pulse variability studies.
     This paper presents a general overview of \psrchive\
     functionality with some focus on the integrated interfaces
     developed for the core applications.}
%%
%     \keywords{pulsars: general --- methods: data analysis}

     \maketitle

%
%________________________________________________ sections below
%
\section{Introduction}

\label{sect:intro}

Within the pulsar astronomy community, a number of individuals and
research groups have developed freely-available software for a wide
variety of purposes.
The {\sc sigproc}\footnote{http://sigproc.sourceforge.net} \citep{lor01b}
and {\sc presto}\footnote{http://www.cv.nrao.edu/~sransom/presto}
\citep{rem02,rce03} software packages are widely used in the search
for new pulsars,
{\sc dspsr}\footnote{http://dspsr.sourceforge.net}
enables real-time phase-coherent dispersion removal \citep{vb11} and
computation of the cyclic spectrum \citep{dem11}, 
both {\sc tempo}\footnote{http://tempo.sourceforge.net} \citep{tw89} and
{\sc tempo2}\footnote{http://www.atnf.csiro.au/research/pulsar/tempo2}
\citep{ehm06} are used in the analysis of pulse arrival time
estimates, 
and {\sc psrcat}\footnote{http://www.atnf.csiro.au/research/pulsar/psrcat}
provides access to the ATNF Pulsar Catalog \citep{mhth05}.
This paper describes some of the basic and advanced functionality of
the \psrchive\footnote{http://psrchive.sourceforge.net} project, which
provides access to a comprehensive range of tools commonly required
for the analysis of pulse profile\footnote{A pulse profile is any
  phase-resolved statistical quantity (e.g. flux density) integrated
  over one or more pulse periods.} data and the various metadata that
describe them \citep{hvm04}.

Each of the above software projects has been refined through stages of
early adoption and beta testing followed by more regular usage and
feedback from the community.
In turn, as the software grows more reliable and reputable, it reduces
barriers to newcomers, thereby promoting growth in the discipline.
These tools are now an indispensable resource to researchers in the
field, and nearly all observational analyses of radio pulsar data
published in the past few decades have relied on one or more of these
packages.

Many of the most fundamental algorithms implemented by \psrchive\
originate in the timing analysis software developed by the
collaborators of the Parkes Southern Pulsar Survey
\citep[PSPS;][]{mld+96}.
Over nearly two decades, these have been generalized, refined and
incorporated into a modular and extensible framework that employs
object-oriented design principles and is primarily implemented using
the C++ language\footnote{http://www.cplusplus.com}.
To increase the portability of the code, it is currently managed using
an open-source distributed version control
system\footnote{http://git-scm.com} and compiled using a
cross-platform build system\footnote{http://sourceware.org/autobook}.

\psrchive\ was developed in parallel with the {\sc
  psrfits}\footnote{http://www.atnf.csiro.au/research/pulsar/psrfits}
file format, which is fully compliant with the Flexible Image
Transport System\footnote{http://fits.gsfc.nasa.gov}
\citep[FITS;][]{hfg+01} endorsed by NASA and the IAU and
compatible with the recommendations of the International Virtual
Observatory Alliance\footnote{http://www.ivoa.net}.
The modular, object-oriented design of the \psrchive\ software
separates the data analysis routines from file I/O, enabling
the software to be easily extended to handle other data formats.
In addition to maintaining backward-compatibility with the file format
used by the original PSPS timing analysis software, {\sc psrchive}
currently provides support to read data in eight different formats,
including the European Pulsar Network flexible format \citep{ljs+98},
the {\sc presto prepfold} output format, and three different file
formats used by pulsar instruments at the Arecibo Observatory.
\psrchive\ automatically determines the format of input data files,
and all of the usage examples presented in this paper can be applied
to any of the supported formats without modification.

The portability and extensibility of \psrchive\ fosters the
incorporation of new features and functionality, including rigorous
polarimetric calibration \citep{van04c,ovhb04}; various methods of
arrival time estimation \citep[e.g.][]{tay92,hbo05,van06}; Faraday
rotation measure determination \citep{hml+06,njkk08}; propagation of
the fourth-order moments of the electric field \citep{van09}; and
statistical analysis of profile variability \citep{dem07,ovh+11}.
Development also continues on some of the more elementary algorithms,
such as estimation of the off-pulse baseline and identification of the
on-pulse region, and computation of the signal-to-noise ratio.

The functionality of \psrchive\ is distributed across a suite of 
specialised programs that are run from the command line in a typical
UNIX shell environment.
A subset of these programs, known as the Core Applications, provide
access to general-purpose routines that are typically required for the
majority of data analyses; these include
\begin{itemize}
\item {\tt psredit}
 - queries or modifies the metadata that describe the data set;
\item {\tt psrstat}
 - derives statistical quantities from the data set and evaluates mathematical expressions;
\item {\tt psrsh}
 - command language interpreter used to transform and reduce data sets; and
\item {\tt psrplot}
 - produces customized, diagnostic and publication-quality plots.
\end{itemize}
The Core Applications employ a standard set of command line options
and incorporate a command language interpreter that can evaluate
mathematical and logical expressions, compute various statistical
quantities, and execute a number of algorithms implemented by 
\psrchive.
Using the Core Applications and data that are available for download
from the CSIRO Data Access Portal, this paper demonstrates a typical
scientific workflow used to analyse observational data and produce
pulse arrival time estimates for high-precision timing.  As part of
this demonstration, \psrchive\ is used to perform radio frequency
interference (RFI) excision, polarimetric calibration, and statistical
bias correction.  Throughout the paper, reference is made to the more
extensive online documentation available at the \psrchive\ web site.
Detailed usage information is also output by each program via the
{\tt -h} command-line option.
In sections~\ref{sect:psredit} through~\ref{sect:psrplot}, each Core
Application is introduced with a description of the motivation and
design of the program followed by a demonstration of its use through a
practical exercise.
Sections~\ref{sect:rfi} through~\ref{sect:timing} demonstrate the use of
\psrchive\ to perform RFI mitigation, polarimetric calibration,
arrival time estimation and bias correction.
The concluding remarks in Section \ref{sect:discussion} include a
description of some \psrchive\ functionality that is currently under
development and some ideas for future work.

\subsection{Observational Data}

The \psrchive\ software processes observational data stored as a
three-dimensional array of pulse profiles; the axes are time
(sub-integration), frequency (channel), and polarization (e.g. the
four Stokes parameters).  The physical properties of the data are
described by various attributes (also called metadata).  A single data
file containing one or more sub-integrations is typically called an
archive.  Sub-integration lengths may be as short as one pulse period
(e.g. for single-pulse studies) or as long as desired (e.g. for the
standard, or template profile, used for high-precision timing).

The examples in this paper make use of nine days of
observations of \psr\ made at 20\,cm with the Parkes 64\,m radio
telescope on 19 to 27 July 2003.
Discovered in the Parkes 70-cm survey \citep{jlh+93}, \psr\
remains the closest and brightest millisecond pulsar known;
it has a spin period of $\sim 5.7$~ms, a pulse width of about 130~$\mu$s
\citep{nms+97} and an average flux of 140~mJy at 20~cm \citep{kxl+98}.
With a sharply-rising main peak and large flux density, it is an
excellent target for high-precision pulsar timing studies.
However, owing to the transition between orthogonally polarized modes
of emission near the peak of the mean pulse profile, arrival time
estimates derived from observations of \psr\ are particularly
sensitive to instrumental calibration errors \citep{sbm+97,van06}.
Furthermore, pulse-to-pulse fluctuations in the emission from this
pulsar on timescales ranging from $\sim10$ to $\sim300$ $\mu$s
\citep{jak+98} place a fundamental limit on the timing precision that
can be achieved \citep{ovh+11}.
These issues are discussed in more detail in
Sections~\ref{sect:calibration} and~\ref{sect:timing}, which
demonstrate the \psrchive\ tools available for mitigating the impact
of polarization calibration errors and correcting the bias due to
self-noise.

As described in the Appendix, these data are available for download
from the CSIRO Data Access Portal using the Pulsar Search tool.
%
% 1 day: on 19 and 20 July 2003
% 10 days: 19 and 27 July 2003
%
A significantly reduced and more readily accessible form of the data
is also available for download from Swinburne University of Technology.
Throughout this paper, it is assumed that the full path to the
directory containing the observational data downloaded from Swinburne
is recorded using the {\tt \$PSRCHIVE\_DATA} shell environment
variable.

\section{Query and modify metadata with {\tt psredit}}
\label{sect:psredit}

The pulse profile data stored in a pulsar archive file are accompanied
by metadata, or attributes, that describe various physical
characteristics of the observation such as the source name, right
ascension and declination, centre radio frequency and bandwidth of the
instrument, start time and duration of the integration, etc.  These
attributes may be queried and modified using the {\tt psredit}
program, which is more fully documented online\footnote{http://psrchive.sourceforge.net/manuals/psredit}.

The keywords used by {\tt psredit} to address the attributes in an
archive are also understood and used by other Core Applications.  For
example, using {\tt psredit} keywords, {\tt psrstat} can perform
variable substitution and evaluate mathematical expressions that
include attribute values; similarly, {\tt psrplot} can annotate plots
(e.g. axes labels and titles) with attribute values as well as any of
the mathematical expressions and statistical quantities provided by
{\tt psrstat}.  This modularity of design allows the interfaces to
the Core Applications to be remembered once and used often. \\ 

\noindent
{\bf Exercise:} In the {\tt \$PSRCHIVE\_DATA/mem} directory, the
receiver name is not set in any of the data files ({\tt *.ar}).  This
can be verified by querying the receiver name attribute.
\begin{Verbatim}[frame=single]
cd $PSRCHIVE_DATA/mem
psredit -c rcvr:name *.ar
\end{Verbatim}
Running {\tt psredit <filename>} with no arguments will print
a listing of every attribute in the file.
Attribute names that end in an asterisk (e.g. {\tt int*:wt*}) represent
vector quantities.  Specifying the attribute name without the asterisk
will print a comma-separated list of every element in the vector; e.g.
to print the centre frequency of every channel in every sub-integration
\begin{Verbatim}[frame=single]
psredit -c int:freq *.ar
\end{Verbatim}
(These data files contain 1 sub-integration and 128 frequency channels.)
To query the value of a single element, or range of elements, a simple
array syntax can be used; e.g.
\begin{Verbatim}[frame=single]
psredit -c 'int:freq[34,56-60]' *.ar
\end{Verbatim}
will print only 6 values for each file.  The single quotation marks in
the above command are necessary to protect the square brackets from
interpretation by the shell. \\

\noindent
Set the receiver name to {\tt MULT\_1} using the {\it standard output
  option}\footnote{http://psrchive.sourceforge.net/manuals/guide/design/options.shtml}
to overwrite the original files.
\begin{Verbatim}[frame=single]
cd $PSRCHIVE_DATA/mem
psredit -c rcvr:name=MULT_1 -m *.ar
\end{Verbatim}
The data files in {\tt \$PSRCHIVE\_DATA/mem/} are a mixture of three different
types.  Create two sub-directories called {\tt pulsar/} and {\tt cal/}
then use {\tt psredit} to query the {\tt type} attribute and use this
to sort the files into the two sub-directories.
\begin{Verbatim}[frame=single,commandchars=\\\(\)]
cd $PSRCHIVE_DATA/mem
mkdir pulsar/
mkdir cal/
mv (\backtick)psredit -c type *.ar | grep Pulsar | awk '{print $1}'\backtick pulsar/
mv *.ar cal/
\end{Verbatim}
The last line of the above commands places all three calibrator
file types in the {\tt cal/} sub-directory.

\section{Evaluate data with {\tt psrstat}}
\label{sect:psrstat}

In addition to accessing the physical attributes that describe the
observation, it is also useful to compute derived quantities, such as
statistical measures that describe the quality of the data, the
effective width of the pulse, the degree of polarisation, etc.
A wide variety of derived quantities can be computed using the {\tt
  psrstat} program, which is more fully documented online\footnote{http://psrchive.sourceforge.net/manuals/psrstat}.
Running {\tt psrstat <filename>} without any command-line
arguments will print a listing of every available quantity.

Any of the quantities (attributes or computed values) provided by the
{\tt psrstat} interface can be substituted into mathematical
expressions that can be evaluated (expressions to be
evaluated are enclosed in braces). For example, to search for
significant peaks in a series of single-pulse archives, query the
maximum amplitude in all phase bins normalized by the off-pulse
standard deviation,
\begin{Verbatim}[frame=single]
psrstat -c '{$all:max/$off:rms}' <filenames>
\end{Verbatim}
To query the effective pulse width in microseconds,
\begin{Verbatim}[frame=single]
psrstat -c '{$weff*$int[0]:period*1e6}' <filename>
\end{Verbatim}

\noindent {\bf Exercise:} Use {\tt psrstat} to print the
signal-to-noise ratio, $S/N$, of each of the pulsar observations.
\begin{Verbatim}[frame=single]
psrstat -c snr pulsar/*.ar
\end{Verbatim}
Note that, by default, {\tt psrstat} computes the $S/N$ of the profile
in the first sub-integration, frequency channel and polarization
(indexed by {\tt subint}, {\tt chan} and {\tt pol}, respectively).
Take one file and print the $S/N$ in each frequency channel using the
command line option to loop over an index.  The number of frequency
channels in the file can be queried with the {\tt nchan} attribute.
\begin{Verbatim}[frame=single]
psredit -c nchan pulsar/n2003200180804.ar
psrstat -l chan=0-127 -c snr pulsar/n2003200180804.ar
\end{Verbatim}
The {\tt psrstat} program can be used in combination with other common
tools (such as {\sc gnuplot}\footnote{http://gnuplot.info}) to investigate problems and/or verify the
quality of data.  When doing so, it is practical to use the {\tt
 -Q} command line option to print only the {\tt value} of each
attribute, instead of {\tt key=value}.

\section{Transform and reduce data with the {\tt psrsh} interpreter}
\label{sect:psrsh}

{\sc psrchive} includes a command language interpreter that provides
access to a large number of common data processing algorithms,
including radio frequency interference mitigation and polarimetric
calibration.  Access to this interpreter is provided by the {\tt psrsh}
program, which may be used either as an interactive shell environment
or as a shell script command processor as more fully
documented online\footnote{http://psrchive.sourceforge.net/manuals/psrsh}.

The {\tt psrsh} interpreter is also embedded in the command line
interfaces of the Core Applications ({\tt psrstat}, {\tt psrplot}, and
{\tt psradd}).  These applications use the interpreter to execute
preprocessing tasks on input data files as described in {\bf Section
  2.3 Job preprocessor} of the online documentation.  The full list of
available commands is listed by running {\tt psrsh -H}.
The first column of the output is the command name, the second column
is a single-letter short-cut key in square brackets, and the third
column is a short description of each command. \\

\noindent
{\bf Exercise:} The $S/N$ values output by {\tt psrstat} in
the previous section are those of only one polarization, not the total
intensity.  The four polarization parameters stored in these files
describe the elements of the coherency matrix: $AA$, $BB$, $\Re[AB]$, and $\Im[AB]$.  The total intensity, $I=AA+BB$ is
formed by the {\tt pscrunch} command.  Use the preprocessing
capability of {\tt psrstat} to print the $S/N$ of the total intensity
as a function of frequency for one file.
\begin{Verbatim}[frame=single]
psrstat -j pscrunch -l chan=0-127 -c snr pulsar/n2003200180804.ar
\end{Verbatim}
Add the {\tt fscrunch} command to print the $S/N$ of the total
intensity integrated across the entire observing bandwidth for each
file.
\begin{Verbatim}[frame=single]
psrstat -j pscrunch,fscrunch -c snr pulsar/*.ar
\end{Verbatim}
Using single-letter short-cut keys, the above line is equivalent to
\begin{Verbatim}[frame=single]
psrstat -j pF -c snr pulsar/*.ar
\end{Verbatim}
The output of the above command (combined with the {\tt -Q} command
line option) can be redirected to a file and {\sc gnuplot} can be used
to plot the variation of $S/N$ as a function of time due to
interstellar scintillation. \\

\section{Display data with {\tt psrplot}}
\label{sect:psrplot}

Using the {\sc pgplot}\footnote{http://www.astro.caltech.edu/~tjp/pgplot}
graphics subroutine library, {\tt psrplot} produces both diagnostic
and publication-quality plots as documented online\footnote{http://psrchive.sourceforge.net/manuals/psrplot}.
The {\tt psrplot} program is a highly configurable plotting tool for
use during all stages of data analysis and manuscript preparation. 
To see a full listing of available plot types, use the {\tt -P}
command-line option.
Each plot can be configured using a wide range of options, including
selection of the range of data to be plotted (zooming), specification
of plotting attributes such as character size and line width, and
definition of plot labels.  Run {\tt psrplot -A <name>} to list the
generic options that are common to most plots, and {\tt psrplot -C
  <name>} to list the options that are specific to the named plot.

Plot labels may include any of the attributes accessible via {\tt
  psredit} and/or quantities and mathematical expressions computed by
{\tt psrstat}.
For example, to produce a publication-quality plot of the
total intensity profile with the $S/N$ printed inside the top-right corner
of the plot frame,
\begin{Verbatim}[frame=single]
psrplot <filename> -pD -jFp -c below:r='S/N: $snr' -c set=pub
\end{Verbatim}
Note that filename(s) need not necessarily be the last argument(s) on
the command line. \\

\noindent
{\bf Exercise:} Use {\tt psrplot} to plot the phase-vs-frequency image of
the total intensity of the pulsar signal in each file.
\begin{Verbatim}[frame=single]
psrplot -p freq -j p pulsar/*.ar
\end{Verbatim}
By default, the dispersion delays between frequency channels are not
corrected.  Using the command line option to loop over an index, plot
the phase-vs-frequency images of $\Re[AB]$ and $\Im[AB]$ ({\tt
  pol=2,3}), which effectively correspond to Stokes $U$ and $V$.
\begin{Verbatim}[frame=single]
psrplot -p freq -l pol=2,3 pulsar/*.ar
\end{Verbatim}
These quantities vary with frequency due to an instrumental effect;
the two orthogonal polarizations propagate through different signal paths
with slightly different lengths, introducing a phase delay that varies
linearly with radio frequency.  \\

\noindent
Use the loop-over-index option to plot the total intensity profile
(the plot type named {\tt flux} or its short-cut {\tt D}) as a
function of frequency in one pulsar data file.
\begin{Verbatim}[frame=single]
psrplot -pD -l chan=0- -jp pulsar/n2003200180804.ar
\end{Verbatim}
Note that {\tt chan=0-} specifies the entire range without having to
know the index of the last frequency channel.  The pulse profile is
significantly distorted at the edges of the band due to quantization
error (also called ``scattered power'') that arises during
analog-to-digital conversion using 2 bits/sample.  The most severely
affected channels will be excised in
Section~\ref{sect:rfi}. \\

\noindent
Plot the Stokes parameters integrated over all frequency channels
for each file. 
\begin{Verbatim}[frame=single]
psrplot -p stokes -jF pulsar/*.ar
\end{Verbatim}
The white line is the total intensity; red, green, and blue correspond
to Stokes $Q$, $U$, and $V$, respectively.
The Stokes parameters vary with time owing to the rotation of the
receiver feed with respect to the sky (the parallactic angle).  In
Section~\ref{sect:calibration}, this effect will be exploited to model
the polarization cross-coupling in the instrumental response and
calibrate the data.

\section{Radio Frequency Interference and Invalid Data Excision}
\label{sect:rfi}

In the typical analysis of observational data, it is necessary to
discard samples that have been corrupted by experimental error,
instrumental distortion, and/or radio frequency interference.  This
section demonstrates some of the {\sc psrchive} algorithms that are
available to assist in the automatic detection and excision of
corrupted data.

\subsection{Excision of frequency channels that are known to be corrupted}
\label{sect:known}

Although the observations of \psr\ used in this paper are
not adversely affected by radio frequency interference, the data near
the edges of the band are known to be corrupted by quantization
distortions. \\

\noindent 
{\bf Exercise:} Use the {\tt psrsh} command language interpreter and
the {\tt zap edge} command to assign zero weight to 15\% of the total bandwidth
($\sim19$ frequency channels) on each edge of the band. Use the {\it output
option} to write output data files with a new extension; e.g. 
\begin{Verbatim}[frame=single]
cd $PSRCHIVE_DATA/mem
psrsh - -e zz pulsar/*.ar cal/*.ar << EOD
zap edge 0.15
EOD
\end{Verbatim}
In the above example, the single hyphen ({\tt -}) command-line option
instructs {\tt psrsh} to read the command script from the standard
input.

\subsection{Automatic detection and excision of narrow-band interference}
\label{sect:median}

Many types of radio frequency interference (RFI) are narrow band, such
that the quality of an observation can be significantly improved by
discarding only a small number of corrupted frequency channels.  The
RFI environment at the telescope may be dynamic, such that it is not
possible to select a fixed set of frequency channels to be excised at
all times.  To address this problem, {\sc psrchive} implements an
automatic frequency channel excision algorithm that is based on
tolerance to differences between the observed spectrum and a version
of the spectrum that has been smoothed by a running median.
By default, the running median is computed using a window that is 21
frequency channels wide and all channels with total flux that differs
from the median-smoothed spectrum by more than 4 times the standard
deviation will be given zero weight.
The standard deviation is defined recursively. That is, the algorithm
works as follows
\begin{enumerate}
\item compute median-smoothed spectrum
\item compute standard deviation, ignoring any zapped channels
\item zap channels that differ from local median by more than tolerance
\item if any channels were zapped, goto 2
\item stop
\end{enumerate}
The above algorithm and its default parameters may not necessarily
work in every situation, and it may require some experimentation to
determine the parameters that detect the majority of RFI for a given
telescope and instrument. \\

\begin{figure}
  \centering
   \includegraphics[angle=-90,width=8cm]{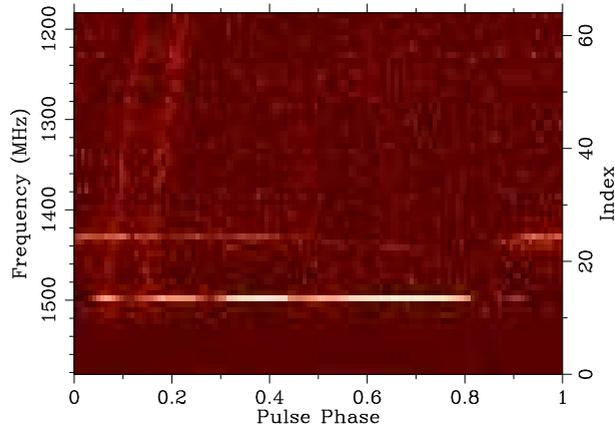}
   \caption{\label{Fig:zap_median}
     Narrow-band radio frequency interference that is not automatically
     detected by the default configuration of the {\tt zap median} algorithm.}
\end{figure}

\begin{figure}
  \centering
   % psrplot -p psd -c set=pub -c x:opt=BCST -c x:lab= -c below:l= example.ar
   \includegraphics[angle=-90,width=8cm]{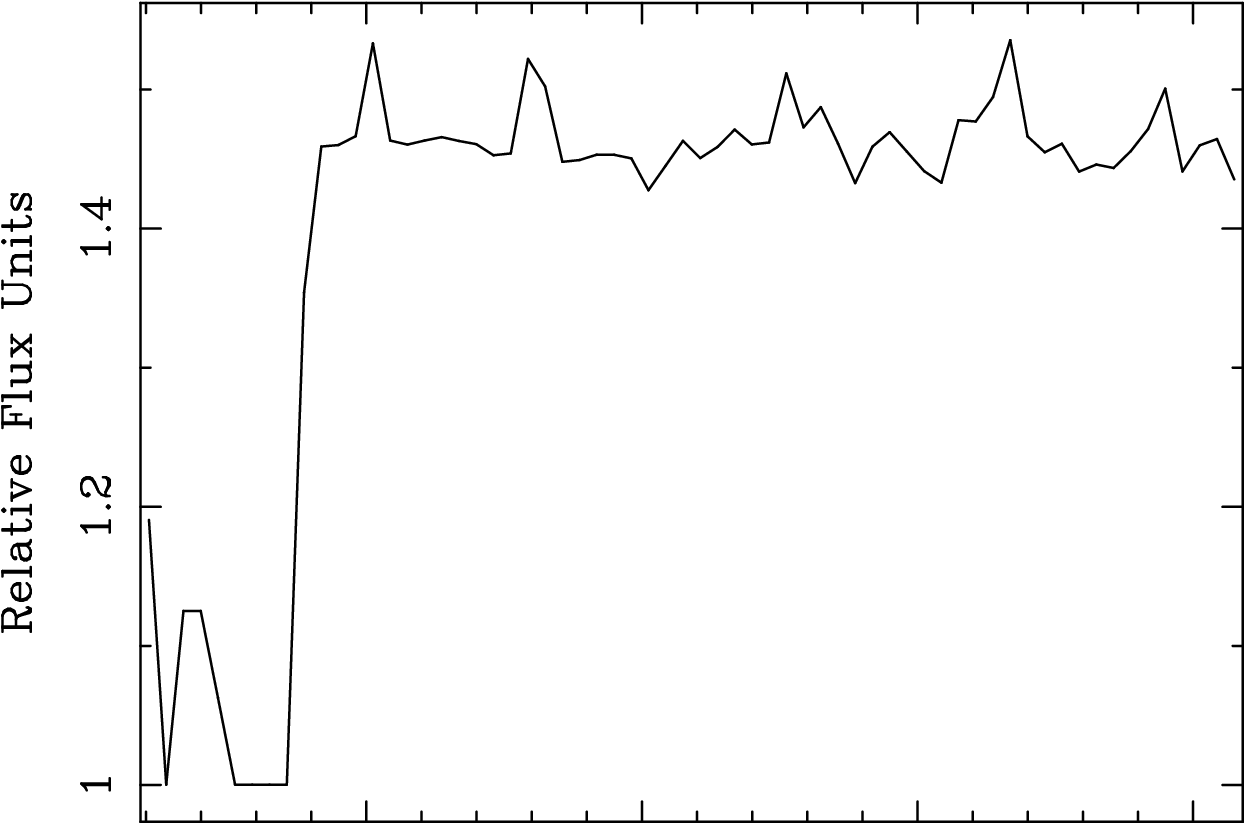}
   \includegraphics[angle=-90,width=8cm]{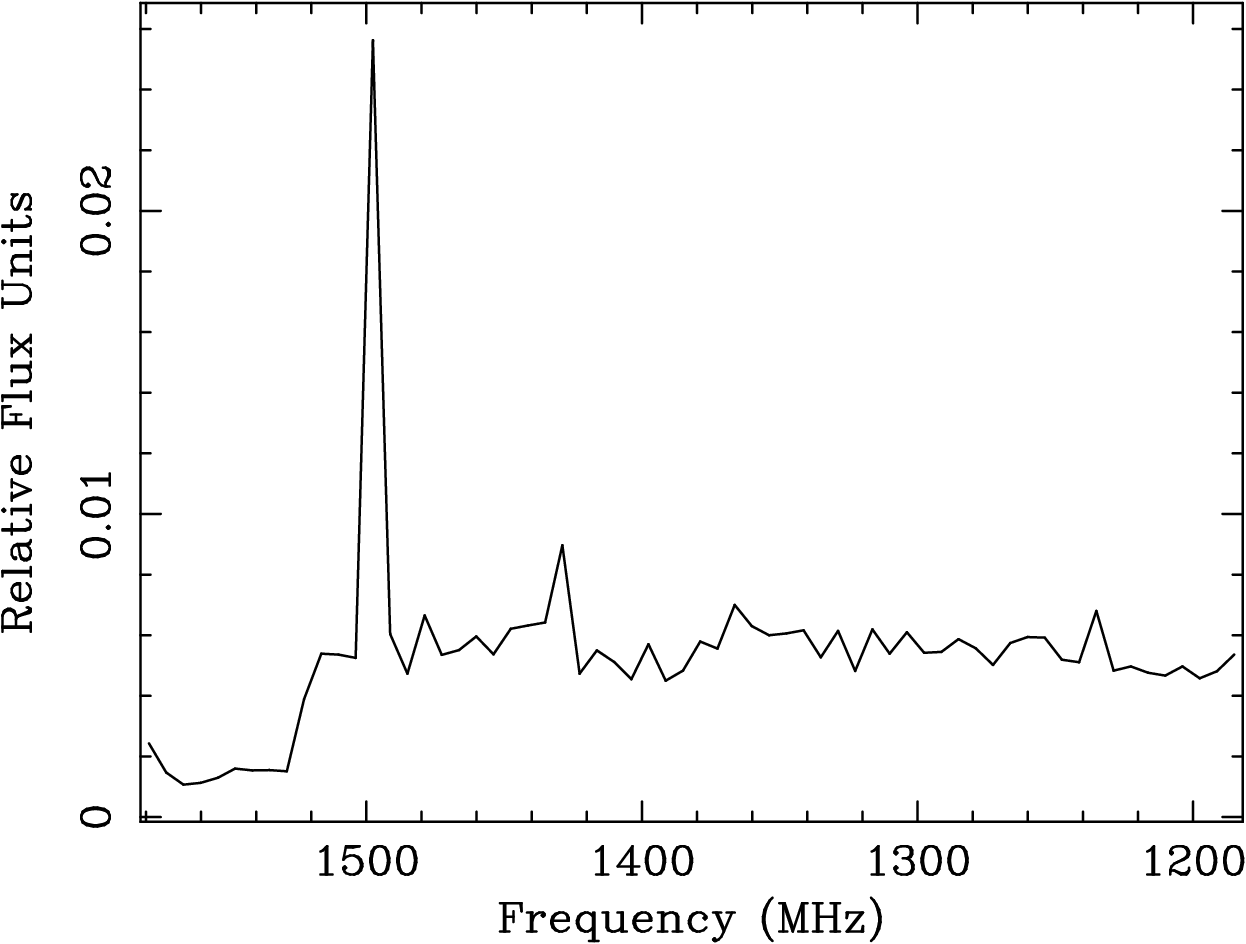}
   \caption{\label{Fig:psd}
     {\it Top panel}: Spectrum formed by summing the values of all pulse phase bins for each frequency channel.
   {\it Bottom panel}: Pulsed spectrum formed by the difference between maximum and minimum values of all pulse phase bins for each frequency channel.  }

\end{figure}

\noindent
{\bf Exercise:} Use {\tt psrplot} and the {\tt zap median}
pre-processing command to view an example of data corrupted by RFI that
is not detected automatically by the default configuration of the
algorithm described above.
\begin{Verbatim}[frame=single]
cd $PSRCHIVE_DATA/zap/BPSR
psrplot -p freq -j "zap median" example.ar
\end{Verbatim}
The plot produced by the above command is shown in
Figure~\ref{Fig:zap_median}.  Note that the signal above 1520 MHz
(below channel index 10) has been filtered prior to digitization.
Narrow-band RFI is evident just below 1430 MHz and 1500 MHz.
To understand why this seemingly-obvious RFI is not detected 
automatically, it is important to note that {\tt psrplot -p freq}
displays the {\it pulsed} flux as a function of pulse phase, whereas
by default {\tt zap median} works with the {\it total} flux summed
over all pulse phases, which is plotted using
\begin{Verbatim}[frame=single]
psrplot -p psd example.ar
\end{Verbatim}
and is shown in the top panel of Figure~\ref{Fig:psd}.
Here, the channels that have been corrupted by RFI are not as obvious.
To utilize a statistic that better characterizes pulsed flux, note
that both the {\tt psd} plot and the {\tt zap median} algorithm can be
configured to use any expression that is understood by {\tt psrstat}.
For example,
\begin{Verbatim}[frame=single]
psrplot -p psd -c 'exp={$all:max-$all:min}' example.ar
\end{Verbatim}
produces the pulsed flux spectrum shown in the bottom panel of
Figure~\ref{Fig:psd}.
The same expression can be used to
configure the {\tt zap median} algorithm, as in the following {\tt psrsh} script

\begin{Verbatim}[frame=lines,label=zap.psh]
#! /usr/bin/env psrsh

# set the expression evaluated in each frequency channel
zap median exp={$all:max-$all:min}

# execute the zap median algorithm
zap median

# zap frequency channels 0 to 8
zap chan 0-8
\end{Verbatim}
This script can be passed to {\tt psrplot} and used to process the
data before plotting.
\begin{Verbatim}[frame=single]
cd $PSRCHIVE_DATA/zap/BPSR
psrplot -p freq -J zap.psh example.ar
\end{Verbatim}
Alternatively, the script can be made executable and run like a
{\sc psrchive} program with the {\it standard output
option} to write the result to a file with a new extension.
\begin{Verbatim}[frame=single]
chmod a+x zap.psh
./zap.psh -e zz example.ar
\end{Verbatim}

\subsection{Automatic detection and removal of impulsive interference}
\label{sect:mow}

Radio frequency interference (RFI) may also occur as broadband bursts
of impulsive emission, such as lightning.  When impulsive interference
is persistent, it may not be possible to discard a subset of corrupted
frequency channels or sub-integrations.  To address this problem, {\sc
  psrchive} implements an automatic impulsive interference mitigation
algorithm that is based on tolerance to differences between the
observed pulse profile and a version of the profile that has been
smoothed by a running median.  To detect impulsive RFI of terrestrial
origin, the pulse profile is first integrated over all frequency
channels without correcting for interstellar dispersion.
The running median is then computed using a window with a duty
cycle of 2\% and any phase bin with total flux that differs
from the median-smoothed profile by more than 4 times the standard
deviation is flagged for replacement.
By default, the standard deviation is defined recursively in a manner
similar to the algorithm used for excising corrupted frequency
channels.
The default recursive standard deviation estimator will fail in
extreme cases of impulsive RFI, and in general it is better to enable
the use of robust statistics (e.g. the median absolute deviation)
as demonstrated in the exercise below.

After flagging corrupted phase bins, the profiles in each frequency
channel and polarization are corrected independently.  A
median-smoothed profile is computed and the values of flagged phase
bins are set equal to the local median plus uncorrupted noise, defined
as the difference between the
observed profile and the local median of a randomly selected phase bin
that has not been flagged as corrupted. \\

\begin{figure}
     \centering
   \includegraphics[angle=-90,width=8cm]{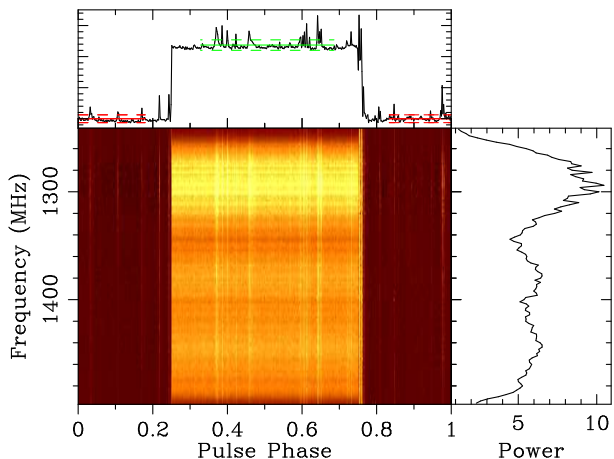}
   \caption{Impulsive radio frequency interference that is not automatically
     detected by the default configuration of the {\tt zap mow} algorithm. }
   \label{Fig:zap_mow}
   \end{figure}

\begin{figure}
     \centering
   \includegraphics[angle=-90,width=8cm]{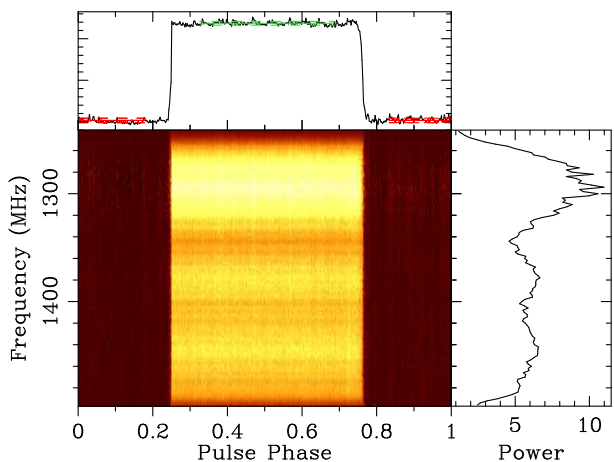}
   \caption{Impulsive radio frequency interference is 
     detected by the {\tt zap mow} algorithm after enabling the use
     of robust statistics and increasing the median smoothing duty
     cycle to 10\%.}
   \label{Fig:zap_mow_robust}
   \end{figure}

\noindent
{\bf Exercise:} Use {\tt psrplot} and the {\tt zap mow}
pre-processing command to view an example of data corrupted by
impulsive RFI that is not detected automatically by the default
configuration of the algorithm.
\begin{Verbatim}[frame=single]
cd $PSRCHIVE_DATA/zap/DFB3
psrplot -p freq+ -jp -j "zap mow" -l subint=0- calibrator.ar
\end{Verbatim}
The above command will produce four separate plots, one for each
sub-integration, the first of which is shown in
Figure~\ref{Fig:zap_mow}.  To enable the use of robust statistics and
increase the median smoothing duty cycle from 2\% to 10\%, create
the following {\tt psrsh} script
\begin{Verbatim}[frame=lines,label=mow.psh]
#! /usr/bin/env psrsh

# use robust statistics (median absolute deviation)
zap mow robust

# set the median-smoothing window to 10% of the pulse profile
zap mow window=0.1

# execute the zap mow algorithm
zap mow
\end{Verbatim}
and pass this script to {\tt psrplot} to process the data before plotting.
\begin{Verbatim}[frame=single]
cd $PSRCHIVE_DATA/zap/DFB3
psrplot -p freq+ -jp -J mow.psh -l subint=0- calibrator.ar
\end{Verbatim}
The first plot produced by running the above command  is shown in
Figure~\ref{Fig:zap_mow_robust}.

\subsection{Interactive excision with {\tt psrzap}}

In some cases, automated methods of detecting corrupted data are
insufficient and it is necessary to perform the task manually.
Large data files with multiple sub-integrations and frequency channels
are best visualized using a dynamic spectrum in which the colour of each
pixel in a two-dimensional image spanned by time and frequency is
determined by a value computed from the pulse profile at that
coordinate.
The {\tt psrzap} program is an interactive tool for excising corrupted
data using the dynamic spectrum.
Run {\tt psrzap -h} for a list of
the keyboard and mouse interactive commands, then 
\begin{Verbatim}[frame=single]
cd $PSRCHIVE_DATA/zap/GUPPI
psrzap guppi_55245_1909-3744_0033_0001.rf
\end{Verbatim}
After loading the data file, two plot windows will open.  The main
window plots the dynamic noise spectrum, which by default is defined
as the variance in each pulse profile as a function of sub-integration
(x-axis) and frequency (y-axis).  The secondary plot window is divided
into three panels: the top panel displays the pulse profile after
integration over all time sub-integrations and frequency channels, the
middle panel displays the phase-versus-frequency image after
integration over all time sub-integrations, and the bottom panel
displays the phase-versus-time image after integration over all
frequency channels.  These diagnostic plots are updated by pressing
{\tt d} on the keyboard.

As with {\tt psrplot -p psd} and {\tt zap median}, the quantity that
is plotted as a function of time and frequency may be specified using
a mathematical expression that is understood by {\tt psrstat}; e.g.
\begin{Verbatim}[frame=single]
psrzap -E '{$all:max-$all:min}' guppi_55245_1909-3744_0033_0001.rf
\end{Verbatim}
There are three modes for selecting ranges of data to view or excise:
\begin{enumerate}
\item time ({\tt t} on keyboard) selects an entire sub-integration (column)
\item frequency ({\tt f} on keyboard) selects an entire frequency channel (row)
\item both ({\tt b} on keyboard) selects a rectangular region
\end{enumerate}
The line(s) passing through the cursor indicate the current selection
mode.  To zoom in on a desired range, click the left mouse button at
the start and end positions of the range.  To excise a desired range,
click the left mouse button at the start, and the right mouse button
at the end of the range.  Simply right clicking will excise the single
column, row, or pixel under the mouse (depending on the selection
mode). \\

\noindent
RFI typically appears as bright spots in the dynamic noise spectrum.
Excise corrupted data until you are satisfied and save the result by
pressing {\tt s} on the keyboard.  
Press {\tt w} to generate a {\tt psrsh} command script that reproduces
the results of the interactive excision session; the script is saved
as a text file with a filename created by appending the {\tt .psh}
extension to the filename of the input archive.  This script can be
integrated into an automated pipeline that reprocesses the original
data from scratch.
Use {\tt psrstat} to compare the $S/N$ of the data before and after
RFI excision.

\begin{Verbatim}[frame=single]
psrstat -jTFp -c snr guppi_55245_1909-3744_0033_0001.rf*
\end{Verbatim}

\section{Polarimetric Calibration}
\label{sect:calibration}

Polarization measurements provide additional insight into the physics
of both the emission and propagation of electromagnetic radiation;
e.g. measurements of Faraday rotation in the interstellar medium yield
constraints on the structure of the Galactic magnetic field
\citep{hml+06} and estimates of the position angle of the linearly
polarized flux indicate that the pulsar spin axis may be aligned with
its space velocity \citep{jhv+05}.
It has also been demonstrated that accurate polarimetry and arrival
time estimation using all four Stokes parameters can signficantly
improve both the accuracy and precision of pulsar timing data
\citep{van06}.

The processes of reception and detection introduce instrumental
artifacts that must be calibrated before meaningful interpretations of
experimental data can be made.
A first-order approximation to calibration can be performed using
observations of a noise diode that is coupled to the receptors, from
which the complex gains of the instrumental response as a function of
frequency are derived. 
This approximation to calibration is based on the ideal feed
assumption that the Jones matrix in each frequency channel has the
form
\begin{equation}
{\bf J}=\left( \begin{array}{cc}
z_0 & 0 \\
0 & z_1
\end{array}\right)
\label{eqn:eigen}
\end{equation}
where $z_0$ and $z_1$ are the complex gains.  The absolute phase of
the Jones matrix is lost during detection, and the matrix may be
parameterized using a polar decomposition described by the absolute
gain $G$, differential gain $\gamma$, and differential phase $\phi$.

\subsection{Display calibrator parameters with {\tt psrplot}}

In \psrchive, calibrator observations of the noise diode have
{\tt type=PolnCal} (as returned by {\tt psredit}); the noise diode
is typically driven by a square wave with a 50\% duty cycle, as
shown in Figures~\ref{Fig:zap_mow} and~\ref{Fig:zap_mow_robust}. \\

\noindent {\bf Exercise:}
Plot the polar decomposition of the ideal feed as a function of frequency.
\begin{Verbatim}[frame=single]
cd $PSRCHIVE_DATA/mem/cal
psrplot -p calm *.zz
\end{Verbatim}

\subsection{Prepare flux calibrator data with {\tt fluxcal}}

\begin{figure}
   \centering
% psrplot -pD -c chan=90 -c set=pub -x -c below:l= n2003201035947.ar
% psrplot -pD -c chan=90 -c set=pub -x -c below:l= n2003201040203.ar -c y:lab=
   \includegraphics[angle=-90,width=8cm]{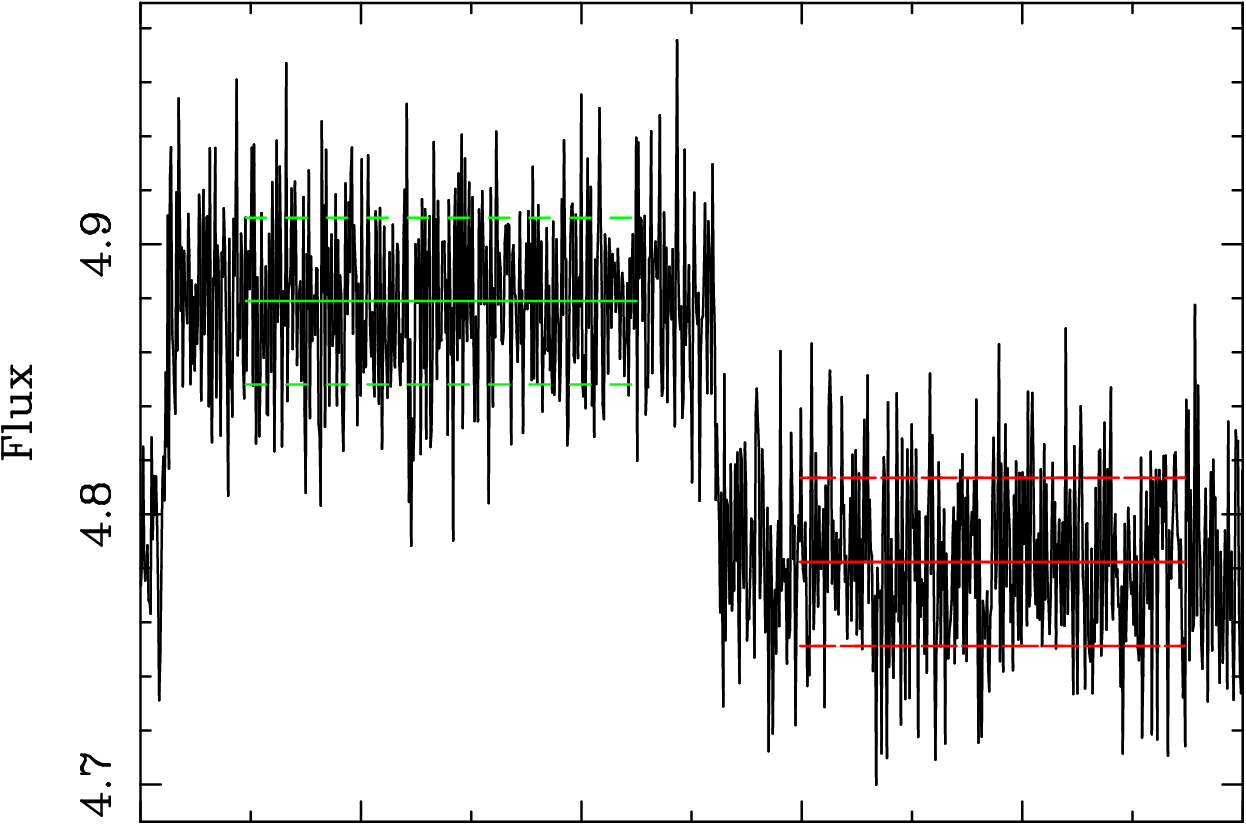}
   \includegraphics[angle=-90,width=8cm]{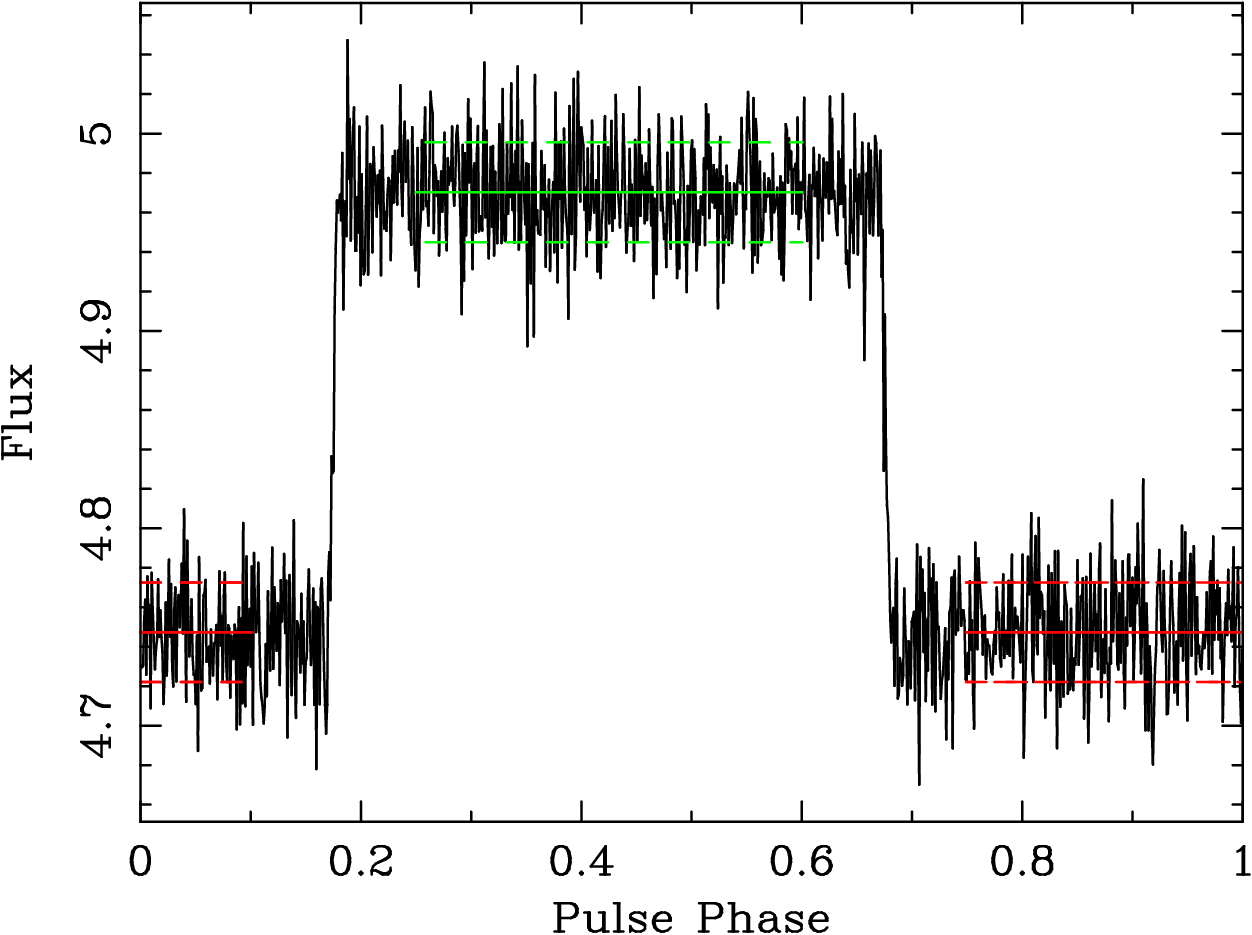}
   \caption{Observations of the Parkes 21-cm Multibeam receiver noise source.
The total intensity from a single 500\,kHz channel was integrated for
approximately 80\,s.  In the top panel, the telescope was pointed at
3C\,218 (Hydra\,A); the mean on-pulse power (green) is used to
estimate \On{H}\ and the mean off-pulse power (red) is used to
estimate \On{L}.  In the bottom panel, the telescope was pointed 2\,deg
north; the mean on-pulse power is used to estimate \Off{H}\ and the
mean off-pulse power is used to estimate \Off{L}. }
\label{Fig:fluxcal}
\end{figure}

Although not immediately necessary for pulsar timing, it is also
useful to perform absolute flux calibration; for example,
well-calibrated estimates of flux density may be used in long-term
studies of refractive scintillation \citep[e.g.][]{rcb84}.
Absolute flux calibration is performed using observations of a {\bf
  standard candle}, an astronomical source with a well-determined
reference flux density and spectral index that applies over a broad
range of radio frequencies.
Two sets of observations are made: the noise diode is driven while the
telescope is pointed at 1) the standard candle and 2) a nearby patch
of sky that is assumed to be empty.  Example plots derived from the
sample data are shown in Figure~\ref{Fig:fluxcal}.
Note that the integration lengths for the on- and off-source
observations need not necessarily be equal (as long as the data
represent mean flux densities). The absolute gains also need not be
equal. Given \\

\begin{tabular}{p{6cm}p{5cm}}
$\On{H} = \On{g} (\Ssys + T_0 + C_0)$   & $\Off{H} = \Off{g} (\Ssys + C_0)$ \\
$\On{L} = \On{g} (\Ssys + T_0)$         & $\Off{L} = \Off{g} \Ssys$ 
\end{tabular} \\

\noindent
where \On{g}\ and \Off{g}\ are the unknown absolute gains of the
instrument while pointing on and off the standard candle, \Ssys\ is
the unknown system equivalent flux density, $S_0$ is the known flux
density of the standard candle, and $C_0$ is the unknown flux density
of the receiver noise source.  Then,
\begin{equation}
\On{f} = {\On{H}\over\On{L}} -1 = {C_0\over\Ssys+S_0}
\end{equation}
and
\begin{equation}
\Off{f} = {\Off{H}\over\Off{L}} -1 = {C_0\over\Ssys}
\label{eqn:Ssys}
\end{equation}
and
\begin{equation}
{1\over\On{f}} - {1\over\Off{f}} = {S_0\over C_0}
\label{eqn:C0}
\end{equation} \\

\noindent
Equation~\ref{eqn:C0} is solved for $C_0$, 
then Equation~\ref{eqn:Ssys} is solved for \Ssys.
In \psrchive, flux calibration observations have a {\tt type}
attribute equal to {\tt FluxCalOn} or {\tt FluxCalOff}.  \\

\noindent
{\bf Exercise:} Start by creating a calibrator database.
\begin{Verbatim}[frame=single]
cd $PSRCHIVE_DATA/mem/cal
pac -w -u zz
\end{Verbatim}
This creates a file called {\tt database.txt}.  Then run
\begin{Verbatim}[frame=single]
fluxcal -f -d database.txt
\end{Verbatim}
This will produce a file named {\tt n2003201035947.fluxcal} and
update {\tt database.txt} with a new entry for this file.
Use {\tt psrplot -p calm} to plot the derived estimates of $\Ssys$ and $S_0$
as a function of radio frequency).

\subsection{Correct the receiver parameters}

A large number of assumptions are built into the design of an
instrument, ranging from the sign of the complex argument in $\exp(\pm i
\omega t)$ to the handedness of circular polarization.  Over time, a
number of inconsitent conventions have been utilised by various
authors; e.g. see \citet{ew01} for a thorough review of contradicting
definitions of the position angle of the linearly polarized flux.

To address this issue, \citet{vmjr10} define the PSR/IEEE convention
that is used by \psrchive\ and include a table of parameters that can
be used to describe the differences between an instrumental design and
the PSR/IEEE convention.  For the Parkes 21-cm Multibeam receiver, it
is necessary to set the symmetry angle to $-\pi/2$, which can be done
with the following command
\begin{Verbatim}[frame=single]
cd $PSRCHIVE_DATA/mem
psredit -c rcvr:sa=-90 -m pulsar/*.ar cal/*.ar
\end{Verbatim}

\subsection{Calibrate using the ideal feed assumption}

To perform the first-order approximation to calibration based on the
ideal feed assumption, run 
\begin{Verbatim}[frame=single]
cd $PSRCHIVE_DATA/mem/pulsar
pac -d ../cal/database.txt *.zz
\end{Verbatim}
For each input file, a new output file will be written with a filename
created by appending the extension {\tt .calib} to the input filename.
If the receiver were ideal, the first-order approximation to
calibration would have eliminated the variation of the Stokes
parameters as a function of parallactic angle.  Use {\tt psrplot} to
test this expectation.
\begin{Verbatim}[frame=single]
psrplot -ps -jF *.calib
\end{Verbatim}
The Stokes parameters still vary as a function of time because the
ideal feed assumption does not apply to the Parkes 21-cm Multibeam
receiver.
Create a sub-directory, e.g. {\tt ideal/} and move the newly
calibrated data to this sub-directory (otherwise, they will be
over-written in the next step).
\begin{Verbatim}[frame=single]
mkdir ideal
mv *.calib ideal/
\end{Verbatim}

\subsection{Measure the cross-coupling parameters using {\tt pcm}}

To accurately calibrate these data, the cross-coupling terms
(off-diagonal components of the Jones matrix) must be estimated, which
can be done by modeling the variation of the Stokes parameters as a
function of time.  The process of performing the least-squares fit is
called Measurement Equation Modeling (MEM); the \psrchive\
implementation is described in \citet{van04c} and more fully
documented online\footnote{http://psrchive.sourceforge.net/manuals/pcm}.
Use {\tt psradd} to combine the archives calibrated using the ideal
feed assumption into a single archive
\begin{Verbatim}[frame=single]
psradd -T -o calib.TT ideal/*.calib
\end{Verbatim}
The resulting archive will be used as input to {\tt pcm}, from
which it will choose the best phase bins to use as constraints and
derive the first guess for the polarization of the source.
%
% As described in {\bf Section 4.1} of the online documentation, run
% {\tt pcm} to derive the cross-coupling parameters.
%
\begin{Verbatim}[frame=single]
pcm -d ../cal/database.txt -s -c calib.TT *.zz
\end{Verbatim}
While running, {\tt pcm} outputs messages about the quality of the
least-squares fits, which are performed independently in each
frequency channel.  On a multi-processor machine, multiple channels
may be solved simultaneously by using the {\tt -t <nthread>}
command-line option, where {\tt <nthread>} is the number of processing
threads to run in parallel.  When {\tt pcm} finishes, it produces an
output file {\tt pcm.fits} that contains the MEM solution; the model
parameters may be plotted using
\begin{Verbatim}[frame=single]
psrplot -p calm pcm.fits
\end{Verbatim}
Compared to the solution derived using the ideal feed assumption,
three new parameters have been added to the model of the receiver: $\theta_1$
describes the non-orthogonality of the feed receptors (the linearly
polarized receptors should be oriented at 0 and 90 degrees) and
$\epsilon_k$ are the ellipticities of the receptors, which should be 0
in an ideal feed with linearly polarized receptors.  The mean value of
$\sim 5$ degrees corresponds to roughly 15\% mixing between linear and
circular polarizations (Stokes $Q$ and $V$).  It is not possible to
determine the absolute rotation of the receptors about the line of
sight, $\theta_0$, without an external reference; therefore, only the
non-orthogonality is measured.

The solution output by {\tt pcm} also includes estimates of the Stokes
parameters of the noise diode, which is no longer assumed to
illuminate both receptors equally and in phase.  This information
enables the calibrator solution derived from one data set to be
applied to observations of another source, as in \citet{ovhb04}.

\subsection{Calibrate using the MEM solution}

Move the output file {\tt pcm.fits} to the {\tt cal/} sub-directory, change
to this directory, and recreate the calibrator database
\begin{Verbatim}[frame=single]
mv pcm.fits ../cal/
cd ../cal
pac -w -u zz -u fits -u fluxcal
\end{Verbatim}
Confirm that {\tt pcm.fits} has been added to {\tt database.txt}, then run
\begin{Verbatim}[frame=single]
cd ../pulsar
pac -d ../cal/database.txt -S *.zz
\end{Verbatim}
Plot the Stokes polarization profile (integrated over the entire band)
in each calibrated data file output by {\tt pac} to confirm that the
Stokes parameters no longer vary with time.

\section{Arrival time estimation}
\label{sect:timing}

In this section, arrival time estimates are derived using a previously
created standard (template) profile.  This high $S/N$ standard profile
was formed by integrating data from $\sim42$ hours of observations.

\subsection{Prepare the standard profile}

The standard profile is located in {\tt \$PSRCHIVE\_DATA/mem/std/standard.ar}.
Plot the phase-vs-frequency image of the total intensity and compare
this image with that of {\tt nonspc.ar} in the same directory.  The
file {\tt nonspc.ar} was formed from data that were not corrected for
scattered power.  Although {\tt standard.ar} was corrected, there are
still residual artefacts in the edges of the band.  Use {\tt psrsh} to
give zero weight to the affected frequency channels then integrate
over all frequency channels.  For best results, excise the same
frequency channels that were excised from the data in
Section~\ref{sect:known}; e.g.
\begin{Verbatim}[frame=single]
cd $PSRCHIVE_DATA/mem/std
psrsh - -e FF standard.ar << EOD
zap edge 0.15
fscrunch
EOD
\end{Verbatim}
If the GNU Scientific
Library\footnote{http://www.gnu.org/software/gsl} is installed and
detected during the configuration of the \psrchive\ software, then it
is possible to use the wavelet smoothing algorithm implemented by
{\tt psrsmooth} to create a ``noise-free'' template profile.
\begin{Verbatim}[frame=single]
psrsmooth -W -t UD8 standard.FF
\end{Verbatim}
This will produce a file called {\tt standard.FF.sm}.  
By default, {\tt psrsmooth} applies the translation-invariant wavelet
denoising algorithm described by \citet{cd95b}.  The profile data are
first transformed into the wavelet domain, where a noise level is
estimated from the data.  Based on the measured noise level, a
threshold is calculated, and all wavelet coefficients with absolute
value below the threshold level are set to zero.  The data are then
transformed back into the profile domain, resulting in a smoothed
profile.
Use the {\tt
crop} attribute of the {\tt flux} plot to zoom in on the low amplitude
flux near the off-pulse baseline and compare the standard profile with
its smoothed version; e.g.
\begin{Verbatim}[frame=single]
psrplot -pD -jp -c crop=0.01 -N 1x2 standard.FF standard.FF.sm
\end{Verbatim}

\subsection{Estimate arrival times using {\tt pat}}

In this section, two different methods of arrival time estimation are
compared: scalar template matching using only the total intensity
\citep{tay92} and matrix template matching using all four Stokes
parameters \citep{van06}.
Comparison is also made between the results derived using
the two different template profiles: the smoothed and
not smoothed versions of {\tt standard.ar}. 
Finally, there are three different data sets: the uncalibrated data,
the data calibrated using the ideal feed assumption, and the data
calibrated using the MEM solution derived with {\tt pcm}.
In total, there are 12 different combinations of arrival time
estimation algorithm, template profile, and observational data.
Experiment with these combinations to find the arrival times with the
lowest residual standard deviation.
To experiment, run {\tt pat} in either scalar
template matching mode; e.g.
\begin{Verbatim}[frame=single]
cd $PSRCHIVE_DATA/mem/pulsar 
pat -F -s ../std/standard.FF *.zz > uncal_unsmooth_stm.tim
\end{Verbatim}
or matrix template matching mode; e.g.
\begin{Verbatim}[frame=single]
pat -Fpc -s ../std/standard.FF.sm *.calib > cal_smooth_mtm.tim
\end{Verbatim}
Run tempo2 to evaluate the arrival times; e.g.
\begin{Verbatim}[frame=single]
tempo2 -f ../pulsar.par uncal_unsmooth_stm.tim
\end{Verbatim}
Search the output of {\tt tempo2} for lines like
\begin{Verbatim}[frame=single]
RMS pre-fit residual = 0.11 (us), RMS post-fit residual = 0.11 (us)
Fit Chisq = 359.7       Chisqr/nfree = 359.74/95 = 3.78671 
\end{Verbatim}
and make note of both the {\bf RMS post-fit residual} and {\bf
Chisqr/nfree} in each case tested.

\subsection{Correct arrival time estimation bias using {\tt psrpca}}
\label{sect:pca}

As the mean flux density of a source of noise approaches the system
equivalent flux density, the statistics of the noise intrinsic to the
source (or self-noise) can no longer be neglected
\citep[e.g.][]{gwi01,van09,gj11}.
This is particularly true when the signal is heavily modulated,
as is typically the case with pulsar emission \citep{ric75},
which can be described as stochastic wide-band impulse modulated 
self-noise \citep[SWIMS;][]{ovh+11}.
Pulsed self-noise is heteroscedastic and, when the timescale of
impulsive modulation is longer than the sampling interval required to
resolve the mean pulse profile, SWIMS is correlated.
Correlated and heteroscedastic noise violates the basic premises of
least-squares estimation and introduces pulse arrival time
measurement bias.
This bias may be corrected using {\tt psrpca}\footnote{The {\tt
    psrpca} program will be compiled only if the GNU Scientific
  Library is installed}, which performs a principle component analysis
of the observed pulse profile shape fluctuations and a multiple
regression analysis in which the post-fit arrival time residual is the
dependent variable and the most significant principal components are
the independent variables.  The methodology is described in detail by
\citet{dem07} and \citet{ovh+11}.

SWIMS is best characterised using a large quantity of data; at the
very least, the number of observations must exceed the number of phase
bins used to resolve the mean pulse profile. This constraint is
satisfied by the example data: $\sim$ 9000 pulse profiles spanning
$\sim 10$ days of observations of \psr\ made at 20 cm with the Parkes
64\,m radio telescope between 19 and 27 July 2003.
To measure and remove the arrival time bias introduced by SWIMS, 
first produce the arrival time estimates
\begin{Verbatim}[frame=single]
cd $PSRCHIVE_DATA/pca
ls -1 *.ar > files.ls
pat -s ../mem/std/standard.FF -f tempo2 -M files.ls > psrpca.tim
\end{Verbatim}
then compute the post-fit arrival time residuals
\begin{Verbatim}[frame=single]
tempo2 -output general2  -s '{sat} {post} {err} SWIMS\n' \
-f ../mem/pulsar.par psrpca.tim | grep SWIMS \
| awk '{print $1,$2,$3}' > resid.dat
\end{Verbatim}
and perform the principal component and multiple regression analyses
\begin{Verbatim}[frame=single]
psrpca -s ../mem/std/stanard.FF -r resid.dat -M files.ls
\end{Verbatim}
When running {\tt psrpca}, it is important to ensure that the arrival
time residuals and input data files are provided in exactly the same
order.  The above commands ensure this by first
creating a file listing named {\tt files.ls}, which is passed to
both {\tt pat} and {\tt psrpca}.
A variety of diagnostic output files are produced by {\tt psrpca},
each with a name that starts with the prefix {\tt psrpca}; this prefix
can be chosen by using the {\tt -p prefix} option.
The diagnostic output files are:
\begin{itemize}
\item {\tt psrpca\_diffs.ar} -- an archive containing the differences
  between the standard template and observed pulse profiles; these
  data are used to construct the covariance matrix;
\item {\tt psrpca\_covariance.dat} -- a plain text file containing the
  covariance matrix;
\item {\tt psrpca\_evals.dat} -- a plain text file containing the
  eigenvalues derived from the covariance matrix in a format that is
  easily inspected using {\sc gnuplot};
\item {\tt psrpca\_evecs.ar} -- a {\sc psrfits} archive containing the
  eigenvectors;
\item {\tt psrpca\_decomposition.dat} -- a plain text file containing
  the decompositions of the observed pulse profiles onto the measured
  eigenvectors;
\item {\tt psrpca\_beta\_zero.dat} and {\tt
    psrpca\_beta\_vector\_used.dat} -- plain text files containing the
  regression coefficients used to remove the bias in arrival time
  estimates; and
\item {\tt psrpca\_residuals.dat} -- a plain text file containing the
  bias-corrected arrival time residuals.
\end{itemize}
These output files enable easy inspection of the principal component
analysis results.
The plain-text output file named {\tt psrpca\_residuals.dat} contains
four columns:
\begin{enumerate}
\item {\tt MJD} -- the site arrival time;
\item {\tt biased\_residual} -- the arrival time residual produced by {\tt tempo2} (expressed in $\mu s$)
\item {\tt corrected\_residual} -- the bias-corrected arrival time residual
\item {\tt error} -- the estimated  measurement error produced by {\tt pat}
\end{enumerate}
This file can be used to compare the biased and corrected arrival time
residuals.
The provided script {\tt rms.sh} may be used to compare the standard
deviation of the two sets of arrival time residuals; e.g.
\begin{Verbatim}[frame=single]
./rms.sh psrpca_residuals.dat 2 4
./rms.sh psrpca_residuals.dat 3 4
\end{Verbatim}
The first command produces the weighted standard deviation and a
measure of goodness of fit for the biased arrival time residuals; the
latter computes the same for the bias-corrected residuals. The biased
and corrected residuals may also be inspected using {\sc gnuplot}.

The primary output of the {\tt psrpca} program is a file named {\tt
  psrpca\_std.ar}.  This {\sc psrfits} archive contains a copy of the
standard template profile that was provided to {\tt psrpca} with an
additional extension that contains the principal component
eigenvectors and the multiple regression coefficients.
{\tt pat} can use the information in this extension to correct the
bias in output arrival time estimates.
With this functionality, the bias predictor derived from one epoch may
be applied to observations made at other epochs and the
bias-corrected arrival time estimates may be provided as input to {\sc
  tempo} or {\sc tempo2}, thereby yielding improved physical parameter
estimates.

\section{Discussion}
\label{sect:discussion}

This paper presents a scientific workflow for high-precision timing
experiments and demonstrates some of the general-purpose tools
applicable to a wider variety of pulsar studies.
In addition to the programs described here, {\sc psrchive} development
continues on a number of novel analysis tools.

For example, the {\tt psrmodel}
program\footnote{http://psrchive.sourceforge.net/manuals/psrmodel}
fits the rotating vector model \citep{rc69a,ew01} to observed
polarisation data using a statistically robust algorithm.
Rather than perform a one-dimensional fit to the real-valued position
angle as a function of pulse phase, {\tt psrmodel}
performs a two-dimensional fit directly to the Stokes $Q$ and Stokes
$U$ profiles by treating them as the real and imaginary components of
a complex number.
For each complex number, the phase is given by the position angle
predicted by the rotating vector model and the magnitude (linearly
polarized flux) is modeled as a free parameter.
This approach has a number of advantages over directly modeling the
position angle: 1) because Stokes $Q$ and $U$ errors are normally
distributed, low $S/N$ data may be included in the fit without biasing
the result; 2) Stokes $Q$ and $U$ are not cyclic; and 3) orthogonal mode
transitions are trivially modeled by negative values of the complex
magnitudes.

Also currently under development, the {\tt psrspa}
program\footnote{http://psrchive.sourceforge.net/manuals/psrspa} can
search for significant single pulses and
derive a wide variety of statistical quantities from single-pulse
data, such as the phase-resolved histogram of the position angle
\citep[e.g.][]{scr+84} and the two-dimensional distribution of the
polarization vector orientation \citep{es04,mck09}. This program
was recently used to analyse the phase distribution and width of
single pulses from the radio magnetar, PSR~J1622$-$4950 \citep{lbb+12}.

A number of the algorithms implemented by \psrchive\ consist of a
single, independent profile transformation that is performed
sequentially by looping over all sub-integrations and frequency
channels.
On a multiprocessor architecture, such transformations could be
readily executed in parallel;
furthermore, identification of such parallelism also provides the
opportunity to apply loop transformations that
improve data locality and conserve memory bandwidth
\citep[e.g.][]{Abu-Sufah:1979:IPV:909095,McKinley:1996:IDL:233561.233564}.

Though written in C++, it is possible to access a large fraction of
{\sc psrchive} functionality via the Python programming
language\footnote{http://www.python.org}.  This provides an
alternative to the {\sc psrchive} applications and {\tt psrsh} command
language interpeter for both high-level scripting of {\sc psrchive}
functionality and interactive or non-standard data manipulation. The
Python interface works by providing direct access to
the core C++ class library on which {\sc psrchive} is built.
Additional information about installing and using the Python interface
is available in the online
documentation\footnote{http://psrchive.sourceforge.net/manuals/python}.
Developers with an interest in extending, refining, or optimising
the \psrchive\ software are encouraged to contact the project administrators
and refer to the extensive online documentation hosted by
{\sc sourceforge}\footnote{http://psrchive.sourceforge.net/devel}. \\

\noindent {\bf Acknowledgements} The authors are grateful to Cees
Bassa, Aidan Hotan, Andrew Jameson, Mike Keith, Jonathan Khoo and Aris
Noutsos for valuable contributions to the {\sc psrchive} software.  We
also thank Han JinLin and his students and colleagues for technical
assistance with the preparation of this manuscript, including
translation of the title and abstract.

\appendix

\section{Obtaining data from the CSIRO Data Access Portal}

To obtain a copy of the observational data used in this paper, visit
the CSIRO Data Access Portal and, using the Pulsar Search tool, enter
the following information
\begin{itemize}
\item Source Name: J0437-4715
\item Project ID: P140
\item Observation Date (dd/mm/yyyy): 19/07/2003 to 27/07/2003
\end{itemize}
%
% On 23 April 2012, this search yielded 266 results.
%
Using the check boxes in the column on the left, refine the search
results to include only ``raw'' observations (i.e. not preprocessed)
with a frequency of 1341\,MHz.  This should yield 67 results, the
first of which has the filename {\tt n2003-07-19-18:08:01.rf}, and a
total download size of 15.3\,GB.

The data stored on the CSIRO Data Access Portal have more time and
frequency resolution than is required for the purposes of the
demonstrations presented in this paper.  Processed versions of these
data with lower resolution are also available for download from
Swinburne University of Technology at
\begin{verbatim}
http://astronomy.swin.edu.au/pulsar/data
\end{verbatim}
Here, there are three files
\begin{itemize}
\item {\tt psrchive\_mem.tgz} (112.6 MB) contains 5-minute
  integrations with full frequency resolution from the first day of
  observations (19 July 2003) -- these data are for use in the
  exercises presented in Sections~\ref{sect:psredit} through
  \ref{sect:timing};
\item {\tt psrchive\_zap.tgz} (299.4 MB) contains three files that are
  not available via the CSIRO Data Access Portal -- these data are for
  use in the exercises presented in Section~\ref{sect:rfi}; and
\item {\tt psrchive\_pca.tgz} (132.4 MB) contains $\sim 17$-second
  integrations with no frequency resolution spanning the nine days of
  observations (from 19 to 27 July 2003) -- these data are for use in
  the exercises presented in Section~\ref{sect:pca}
\end{itemize}
These files can be unpacked with commands such as
\begin{Verbatim}[frame=single]
gunzip -c psrchive_mem.tgz | tar xf -
\end{Verbatim}
The examples presented throughout this paper assume that the above
three files have been downloaded from Swinburne University of
Technology and the full path to the directory into which they have
been unpacked is recorded using the {\tt \$PSRCHIVE\_DATA} shell
environment variable.

\bibliographystyle{raa}
%\bibliography{journals,modrefs,psrrefs,../local,crossrefs}

\begin{thebibliography}{42}
\providecommand{\natexlab}[1]{#1}
\providecommand{\selectlanguage}[1]{\relax}

\bibitem[{Abu-Sufah(1979)}]{Abu-Sufah:1979:IPV:909095}
Abu-Sufah, W. A.-K. 1979, Improving the performance of virtual memory
  computers., Ph.D. thesis, Champaign, IL, USA

\bibitem[{Coifman \& Donoho(1995)}]{cd95b}
Coifman, R.~R., \& Donoho, D.~L. 1995, in Wavelets and Statistics, Springer
  Lecture Notes in Statistics, vol. 103, 125--150

\bibitem[{Demorest(2007)}]{dem07}
Demorest, P.~B. 2007, Measuring the Gravitational Wave Background using
  Precision Pulsar Timing, Ph.D. thesis, University of California, Berkeley

\bibitem[{{Demorest}(2011)}]{dem11}
{Demorest}, P.~B. 2011, MNRAS, 416, 2821

\bibitem[{{Edwards} et~al.(2006){Edwards}, {Hobbs}, \& {Manchester}}]{ehm06}
{Edwards}, R.~T., {Hobbs}, G.~B., \& {Manchester}, R.~N. 2006, MNRAS, 372, 1549

\bibitem[{{Edwards} \& {Stappers}(2004)}]{es04}
{Edwards}, R.~T., \& {Stappers}, B.~W. 2004, A\&A, 421, 681

\bibitem[{{Everett} \& {Weisberg}(2001)}]{ew01}
{Everett}, J.~E., \& {Weisberg}, J.~M. 2001, ApJ, 553, 341

\bibitem[{Gwinn(2001)}]{gwi01}
Gwinn, C.~R. 2001, ApJ, 561, 815

\bibitem[{{Gwinn} \& {Johnson}(2011)}]{gj11}
{Gwinn}, C.~R., \& {Johnson}, M.~D. 2011, ApJ, 733, 51

\bibitem[{Han et~al.(2006)Han, Manchester, Lyne, Qiao, \& van Straten}]{hml+06}
Han, J.~L., Manchester, R.~N., Lyne, A.~G., Qiao, G.~J., \& van Straten, W.
  2006, ApJ, 642, 868

\bibitem[{{Hanisch} et~al.(2001){Hanisch}, {Farris}, {Greisen} et~al.}]{hfg+01}
{Hanisch}, R.~J., {Farris}, A., {Greisen}, E.~W., et~al. 2001, A\&A, 376, 359

\bibitem[{{Hotan} et~al.(2005){Hotan}, {Bailes}, \& {Ord}}]{hbo05}
{Hotan}, A.~W., {Bailes}, M., \& {Ord}, S.~M. 2005, ApJ, 624, 906

\bibitem[{{Hotan} et~al.(2004){Hotan}, {van Straten}, \& {Manchester}}]{hvm04}
{Hotan}, A.~W., {van Straten}, W., \& {Manchester}, R.~N. 2004, PASA, 21, 302

\bibitem[{Jenet et~al.(1998)Jenet, Anderson, Kaspi, Prince, \& Unwin}]{jak+98}
Jenet, F., Anderson, S., Kaspi, V., Prince, T., \& Unwin, S. 1998, ApJ, 498,
  365

\bibitem[{{Johnston} et~al.(2005){Johnston}, {Hobbs}, {Vigeland}
  et~al.}]{jhv+05}
{Johnston}, S., {Hobbs}, G., {Vigeland}, S., et~al. 2005, MNRAS, 364, 1397

\bibitem[{Johnston et~al.(1993)Johnston, Lorimer, Harrison et~al.}]{jlh+93}
Johnston, S., Lorimer, D.~R., Harrison, P.~A., et~al. 1993, Nature, 361, 613

\bibitem[{Kramer et~al.(1998)Kramer, Xilouris, Lorimer et~al.}]{kxl+98}
Kramer, M., Xilouris, K.~M., Lorimer, D.~R., et~al. 1998, ApJ, 501, 270

\bibitem[{{Levin} et~al.(2012){Levin}, {Bailes}, {Bates} et~al.}]{lbb+12}
{Levin}, L., {Bailes}, M., {Bates}, S.~D., et~al. 2012, MNRAS, 2751

\bibitem[{{Lorimer}(2001)}]{lor01b}
{Lorimer}, D.~R. 2001, \emph{{SIGPROC-v1.0: (Pulsar) Signal Processing
  Programs}}, {Arecibo Technical Memo No.~2001--01}

\bibitem[{Lorimer et~al.(1998)Lorimer, Jessner, Seiradakis et~al.}]{ljs+98}
Lorimer, D.~R., Jessner, A., Seiradakis, J.~H., et~al. 1998, A\&AS, 128, 541

\bibitem[{{Manchester} et~al.(2005){Manchester}, {Hobbs}, {Teoh}, \&
  {Hobbs}}]{mhth05}
{Manchester}, R.~N., {Hobbs}, G.~B., {Teoh}, A., \& {Hobbs}, M. 2005, AJ, 129,
  1993

\bibitem[{Manchester et~al.(1996)Manchester, Lyne, D'Amico et~al.}]{mld+96}
Manchester, R.~N., Lyne, A.~G., D'Amico, N., et~al. 1996, MNRAS, 279, 1235

\bibitem[{McKinley et~al.(1996)McKinley, Carr, \&
  Tseng}]{McKinley:1996:IDL:233561.233564}
McKinley, K.~S., Carr, S., \& Tseng, C.-W. 1996, ACM Trans. Program. Lang.
  Syst., 18, 424

\bibitem[{{McKinnon}(2009)}]{mck09}
{McKinnon}, M.~M. 2009, ApJ, 692, 459

\bibitem[{Navarro et~al.(1997)Navarro, Manchester, Sandhu, Kulkarni, \&
  Bailes}]{nms+97}
Navarro, J., Manchester, R.~N., Sandhu, J.~S., Kulkarni, S.~R., \& Bailes, M.
  1997, ApJ, 486, 1019

\bibitem[{{Noutsos} et~al.(2008){Noutsos}, {Johnston}, {Kramer}, \&
  {Karastergiou}}]{njkk08}
{Noutsos}, A., {Johnston}, S., {Kramer}, M., \& {Karastergiou}, A. 2008, MNRAS,
  386, 1881

\bibitem[{{Ord} et~al.(2004){Ord}, {van Straten}, {Hotan}, \&
  {Bailes}}]{ovhb04}
{Ord}, S.~M., {van Straten}, W., {Hotan}, A.~W., \& {Bailes}, M. 2004, MNRAS,
  352, 804

\bibitem[{{Os{\l}owski} et~al.(2011){Os{\l}owski}, {van Straten}, {Hobbs},
  {Bailes}, \& {Demorest}}]{ovh+11}
{Os{\l}owski}, S., {van Straten}, W., {Hobbs}, G.~B., {Bailes}, M., \&
  {Demorest}, P. 2011, MNRAS, 418, 1258

\bibitem[{Radhakrishnan \& Cooke(1969)}]{rc69a}
Radhakrishnan, V., \& Cooke, D.~J. 1969, Astrophys. Lett., 3, 225

\bibitem[{{Ransom} et~al.(2003){Ransom}, {Cordes}, \& {Eikenberry}}]{rce03}
{Ransom}, S.~M., {Cordes}, J.~M., \& {Eikenberry}, S.~S. 2003, ApJ, 589, 911

\bibitem[{{Ransom} et~al.(2002){Ransom}, {Eikenberry}, \&
  {Middleditch}}]{rem02}
{Ransom}, S.~M., {Eikenberry}, S.~S., \& {Middleditch}, J. 2002, AJ, 124, 1788

\bibitem[{{Rickett}(1975)}]{ric75}
{Rickett}, B.~J. 1975, ApJ, 197, 185

\bibitem[{Rickett et~al.(1984)Rickett, Coles, \& Bourgois}]{rcb84}
Rickett, B.~J., Coles, W.~A., \& Bourgois, G. 1984, A\&A, 134, 390

\bibitem[{Sandhu et~al.(1997)Sandhu, Bailes, Manchester et~al.}]{sbm+97}
Sandhu, J.~S., Bailes, M., Manchester, R.~N., et~al. 1997, ApJ, 478, L95

\bibitem[{Stinebring et~al.(1984)Stinebring, Cordes, Rankin, Weisberg, \&
  Boriakoff}]{scr+84}
Stinebring, D.~R., Cordes, J.~M., Rankin, J.~M., Weisberg, J.~M., \& Boriakoff,
  V. 1984, ApJS, 55, 247

\bibitem[{Taylor(1992)}]{tay92}
Taylor, J.~H. 1992, Philos. Trans. Roy. Soc. London A, 341, 117

\bibitem[{Taylor \& Weisberg(1989)}]{tw89}
Taylor, J.~H., \& Weisberg, J.~M. 1989, ApJ, 345, 434

\bibitem[{{van Straten}(2004)}]{van04c}
{van Straten}, W. 2004, ApJS, 152, 129

\bibitem[{{van Straten}(2006)}]{van06}
{van Straten}, W. 2006, ApJ, 642, 1004

\bibitem[{{van Straten}(2009)}]{van09}
{van Straten}, W. 2009, ApJ, 694, 1413

\bibitem[{{van Straten} \& {Bailes}(2011)}]{vb11}
{van Straten}, W., \& {Bailes}, M. 2011, PASA, 28, 1

\bibitem[{{van Straten} et~al.(2010){van Straten}, {Manchester}, {Johnston}, \&
  {Reynolds}}]{vmjr10}
{van Straten}, W., {Manchester}, R.~N., {Johnston}, S., \& {Reynolds}, J.~E.
  2010, PASA, 27, 104

\end{thebibliography}

\label{lastpage}
\end{document}